\documentclass[prb,aps,twocolumn,showpacs]{revtex4}
\usepackage{epsfig}
\begin{document}

\title{Electronic Structure of Electron-doped Sm$_{1.86}$Ce$_{0.14}$CuO$_4$:
Strong `Pseudo-Gap' Effects, Nodeless Gap and Signatures of Short
Range Order}

\author{S. R. Park$^1$, Y. S. Roh$^1$, Y. K. Yoon$^1$, C. S. Leem$^1$,
J. H. Kim$^1$, B. J. Kim$^2$, H. Koh$^3$, H. Eisaki$^4$, N. P.
Armitage$^{5,6}$, C. Kim$^1$}

\affiliation{$^1$Institute of Physics and Applied Physics, Yonsei
University, Seoul, Korea}

\affiliation{$^2$School of Physics and Center for Strongly
Correlated Materials Research, Seoul National University, Seoul,
Korea}

\affiliation{$^3$Advanced Light Source, Lawrence Berkeley National
Laboratory, Berkeley, California 94720, USA}

\affiliation{$^4$Advanced Industrial Science and Technology,
Tsukuba, Japan}

\affiliation{$^5$D\'{e}partement de Physique de la Mati\`{e}re
Condens\'{e}e, Universit\'{e} de Gen\`{e}ve, quai Ernest-Ansermet
24, CH1211 Gen\`{e}ve 4, Switzerland}

\affiliation{$^6$Department of Physics and Astronomy, The Johns
Hopkins University, Baltimore, MD 21218}

\date{\today}

\begin{abstract}

Angle resolved photoemission (ARPES) data from the electron doped
cuprate superconductor Sm$_{1.86}$Ce$_{0.14}$CuO$_4$ shows a much
stronger pseudo-gap or ``hot-spot" effect than that observed in
other optimally doped $n$-type cuprates. Importantly, these effects
are strong enough to drive the zone-diagonal states below the
chemical potential, implying that $d$-wave superconductivity in this
compound would be of a novel ``nodeless" gap variety. The gross
features of the Fermi surface topology and low energy electronic
structure are found to be well described by reconstruction of bands
by a $\sqrt{2}\times\sqrt{2}$ order. Comparison of the ARPES and
optical data from the $same$ sample shows that the pseudo-gap energy
observed in optical data is consistent with the inter-band
transition energy of the model, allowing us to have a unified
picture of pseudo-gap effects. However, the high energy electronic
structure is found to be inconsistent with such a scenario. We show
that a number of these model inconsistencies can be resolved by
considering a short range ordering or inhomogeneous state.
\end{abstract}
\maketitle

In spite of their many interesting physical
properties\cite{ZWang,Dagan,Matsuda,Yamada}, electron-doped HTSCs
have been much less studied as compared to the hole-doped HTSCs. It
was not until a few years ago that high resolution ARPES was applied
and there are still only a handful of published papers on the
subject. In their high resolution ARPES studies, Armitage {\it et
al.} found that for the highest T$_c$ samples of
Nd$_{1.85}$Ce$_{0.15}$CuO$_4$ (NCCO) the near E$_F$ spectral weight
was strongly suppressed at the momentum space positions where the
underlying Fermi surface (FS) contour crosses the AF Brillouin zone
boundary (AFZB) suggesting the existence of a ($\pi$,$\pi$)
scattering channel \cite{Armitage2}. It was also found that for
(x=0.04) underdoped samples an electron FS pocket exists around
($\pi$,0) point and that at higher dopings spectral weight increases
near ($\pi/2$,$\pi/2$) which eventually completes the large
hole-like FS pocket around the ($\pi$,$\pi$) point \cite{Armitage3}.

A possible way to view the results for the highest-T$_c$ samples -
at least qualitatively - is as a manifestation of a band
reconstruction from a $\sqrt{2}\times\sqrt{2}$ static (or slowly
fluctuating) spin density wave (SDW) or similar symmetry order
\cite{NPAthesis}. Such a picture explains hot spots on the FS
contour as due not to the opening of a `pseudo-gap' $per$ $se$ but
instead due to a band folding and then splitting across the AFZB
giving FS pockets around ($\pi/2,\pi/2$) and ($\pi,0$). Such a
simple two band interpretation enables one to understand issues
such as the sign change in the Hall coefficient\cite{MillisHall}
and optical conductivity \cite{Zimmers} spectra. However,
systematic studies to test the model are lacking and there may be
doubts that such a simple picture could describe the data at the
level of small details.

Motivated by these issues, we have performed an extensive high
resolution ARPES study on another compound in the small family of
electron-doped HTSCs Sm$_{1.85}$Ce$_{0.15}$CuO$_4$ (SCCO) as well
as optical reflection measurements.  A quantitative analysis
yields a number of important observations, which have implications
for the gap symmetry and pairing mechanism.  We find a much
stronger pseudo-gap or ``hot-spot" effect that creates a gap where
a node is expected in a $d$-wave superconductor.  Additionally, we
find that - despite its simplicity - a SDW scenario\cite{SDW} can
account for various aspects of the low energy electronic structure
quite well, such as the Fermi surface topology and low energy band
dispersions. The data also clearly show the link between the
optical pseudo-gap and the inter-band transition between the split
bands. We argue that other anomalous aspects may be explained
through the consideration of a short-range ordering or
inhomogeneous state.

SCCO single crystal were grown by the travelling-solvent
floating-zone method. The doping level determined by the inductive
coupled plasma method was $x=0.14$. As grown samples (unreduced)
are not superconducting due to the believe presence of extra
oxygen at the apical sites. Crystals were `reduced' by annealing
in Ar for 48 hours at 1000 C, and then in oxygen for 24 hours at
500 C to induce superconductivity. T$_c$ was determined to be 13 K
by magnetic susceptibility measurement. ARPES experiments were
performed at beamline 7.0.1 at the Advanced Light Source. 85 and
135 eV photon energies were used with a total instrumental energy
and angular resolutions of approximately 50 meV and 0.4$^{\circ}$,
respectively. Additional higher resolution data were taken at
beamline 5-4 of the Stanford Synchrotron Radiation Laboratory
using 16.5 eV photons with an energy resolution of 10 meV. Samples
were cleaved {\it in situ} and laser aligned. Superior surface
quality upon cleaving and small lineshape differences allow us to
resolve features that were not observed on related compounds. The
chamber pressure was better than $4\times 10^{-11}$ torr and the
temperature was kept between 15 and 20 K.

\begin{figure}[t]
\centering \leavevmode \epsfxsize=8.0cm \epsfbox{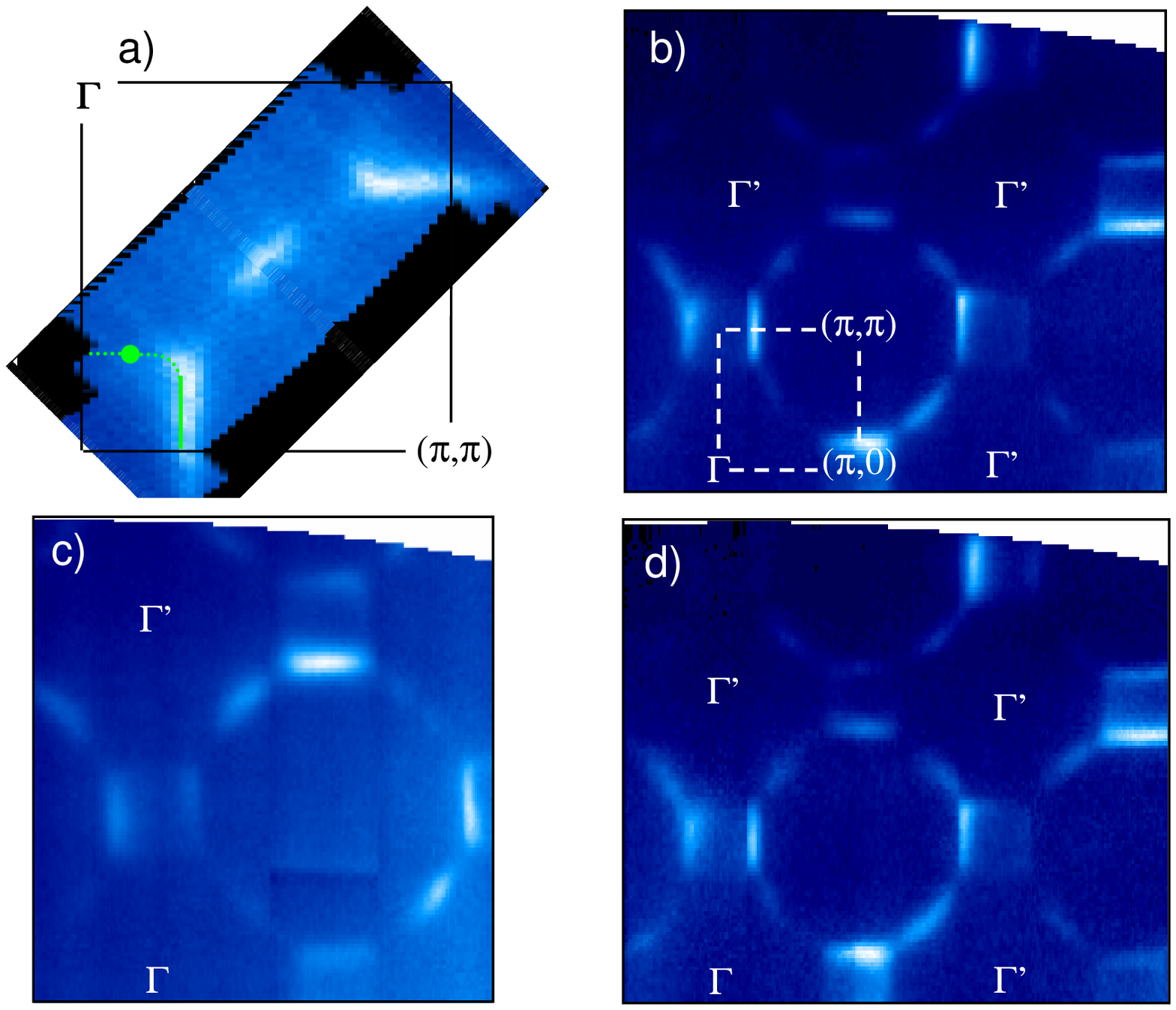}
\vspace{.0cm} \caption{Fermi surface maps of (a) a reduced sample
taken with 16.5 eV photons, (b) a reduced sample at the Sm
4$d\rightarrow$ 4$f$ edge (135 eV), and (c) an unreduced sample with
85 eV. (d) Momentum distribution plot at 100 meV binding energy. In
(b), the first quadrant of the first Brillouin zone is drawn as a
dotted square.}\label{fig1}
\end{figure}

In Fig. 1a, we show a spectral intensity map obtained by integrating
the spectra within a 30 meV window at the Fermi energy. The photon
energy is 16.5 eV. The high intensity locus in momentum space can be
interpreted as the FS. One immediately notices that the FS contour
has a suppression of the spectral intensity where it crosses the
AFZB (($\pi$,0)-(0,$\pi$) line) as observed previously for
NCCO\cite{Armitage2}. Even though such a segmented FS has been
observed for other compounds which suggests that it is an intrinsic
property of the CuO planes in the electron doped HTSCs, we performed
additional experiments to make sure that all the FS segments are CuO
states derived and that our observation is not due to photoemission
matrix element effects.

The FS map shown in panel (b) was taken with the photon energy
tuned to the Sm 4$d\rightarrow$ 4$f$ edge (135 eV) and looks
similar to the map taken with 16.5 eV photons. This excludes the
possibility that any of the broken FS segments are related to Sm
derived states as its intensity would have been relatively
enhanced due to the resonance effect. In addition, the FS
suppression is seen in higher Brillouin zones (BZ) in which the
photoemission selection rules are different from that in the first
BZ. Shown in panel (c) is a similar map taken on an unreduced
(non-superconducting) sample with 85 eV photons. Even though we
were not able to obtain detailed information due to the lower
resolution, the gross features of the spectra from unreduced
samples are qualitatively similar to that of reduced samples. The
above results confirm that the FS suppression is an intrinsic
property of CuO planes in electron doped HTSCs.

One can obtain more information by looking at the data at binding
energies higher than E$_F$. Fig. 1d shows the momentum
distribution of the spectral intensity at 100 meV binding energy
instead of at E$_F$. One finds that the `break' in the momentum
distribution still exists at this energy and therefore the
spectral weight suppression does not derive from the opening of a
low energy pseudo-gap at E$_F$ but suggests that it involves
energy scales of at least 100 meV. This in turn points to the
scattering being from {\it static} (or quasi-static)
origin\cite{Onufrieva}.

We therefore analyze our data based on a simple
$\sqrt{2}\times\sqrt{2}$ SDW model band structure with
characteristic wave vector ($\pi$,$\pi$). We note that although an
SDW is the natural choice based on the close proximity of the
antiferromagnetic phase, our data is consistent with any ordering
of characteristic wave vector ($\pi$,$\pi$).  The
$\sqrt{2}\times\sqrt{2}$ band structure can be obtained via simple
degenerate perturbation theory by solving a $2\times 2$ matrix,
which gives
\begin{eqnarray}
E_k=E_0+4t'(\cos{k_x}\cos{k_y})+2t''(\cos{2k_x}+\cos{2k_y}) \nonumber \\
\pm\sqrt{4t^2(\cos{k_x}+\cos{k_y})^2+|V_{\pi\pi}|^2}
\end{eqnarray}
where $V_{\pi\pi}$ is the strength of the effective ($\pi$,$\pi$)
scattering, and $t$, $t'$ and $t''$ are hopping energies. The
original large FS band centered around ($\pi,\pi$) characterized by
the three hopping energies $t$, $t'$ and $t''$ reconstructs due to
the ($\pi$,$\pi$) scattering and splits into two bands. We analyze
the data quantitatively within this model.

\begin{figure}[t]
\centering \leavevmode \epsfxsize=8.5cm \epsfbox{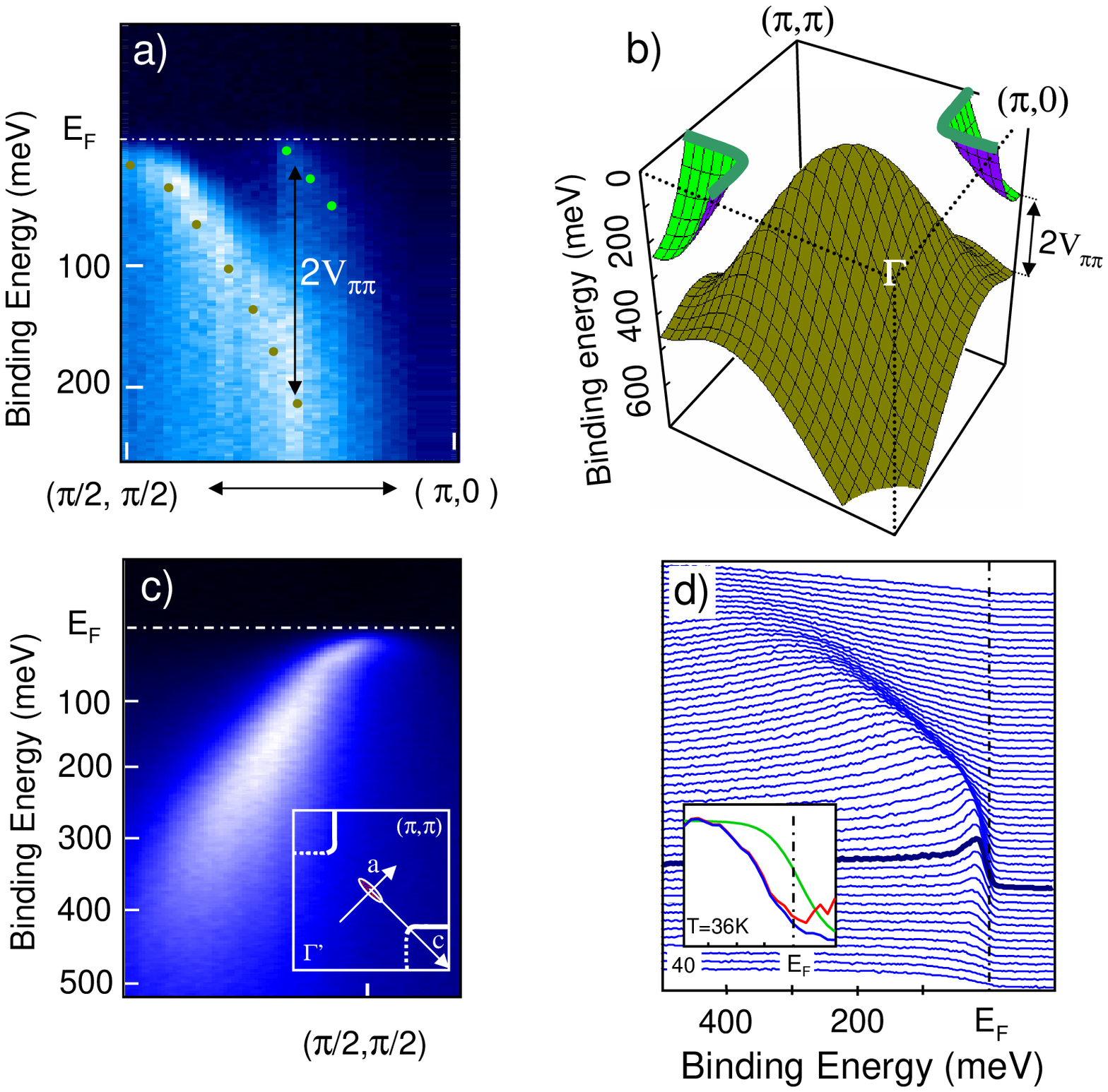}
\vspace{.0cm} \caption{(a) Cut along the AFZB. (b) Schematic of
the reconstructed band structure in the model. (c) Data along the
$\Gamma$-($\pi$,$\pi$) cut. Inset shows the FS as well as the
location where data in (a) and (c) are taken. (d) EDCs of the data
in (c). The inset shows the EDC at the E$_F$ crossing (blue), Au
spectrum (green) and the EDC divided by the latter
(red).}\label{fig2}
\end{figure}

The best way to view the band splitting is to look at the data along
the AFZB as the splitting along this cut should show a constant
value of 2$V_{\pi\pi}$ within the model. Panel (a) in Fig. 2 shows
the data along the AFZB taken with 16.5 eV photons. One can see two
bands with a roughly constant splitting, consistent with the model
and theoretical results\cite{Tohyama}. We note that this observation
clearly contradicts the recent claim of anisotropic spin correlation
gap\cite{Matsui2}. One also sees that unlike recent data on NCCO
there are no quasiparticle-like peaks or even FS crossing in the hot
spot itself \cite{Matsui1}. Fitting of the experimental dispersion
in panel (a) as well as the FSs in Fig. 1 within the model gives us
$t$=0.25 eV, $t'$=0.036 eV, $t''$=0.027 eV and $V_{\pi\pi}$=0.11 eV.
The band structure constructed with the extracted parameters is
depicted in panel (b). We see that the lower one of the split bands
may contribute to a small hole-like FS pocket near ($\pi/2$,$\pi/2$)
if it crosses E$_F$ while the upper one gives a electron-like FS
centered at ($\pi$,0). We note that the observed $V_{\pi\pi}$ value
is larger than the analogous one in NCCO\cite{Armitage2}.

As shown in Fig. 2b, although the model predicts band folding should
occur across the AFZB, such folding has not been observed in the
published data to date. Our new data along the $\Gamma$ to
($\pi$,$\pi$) cut shown in panel (c) shows such a band folding as we
go from $\Gamma$ to ($\pi$,$\pi$), the band approaches E$_F$, bends
at around 50 meV and appears to fold back near ($\pi/2$,$\pi/2$). It
is clearly seen in the energy distribution curves (EDCs) of the same
data in panel (d). We attribute the observance of band folding in
SCCO to the better quality of the data and a larger $V_{\pi\pi}$.
The quasi-particle-like sharp peak was comparatively smaller in the
case of NCCO with a width approximately double\cite{Armitage4}. This
appears to indicate a better surface quality for SCCO.

The $\sqrt{2}\times\sqrt{2}$ band model also allows one to explain
the peculiar straight section of the FS near ($\pi,0$) as deriving
from the original curved FS contour being straightened due to the
band reconstruction. This would be very hard to understand if the
break in the FS contour were due to a simple suppression of the FS
in a single band. In addition, fitting of the FS by a single
contour and counting the electrons in the occupied $k$-space
region gives a FS volume of $1.07\pm 0.01$ which is smaller than
the expected value of 1.14 (14$\%$ doping). Whereas, the FS volume
measured based on the two band model depicted in Fig. 2b is
$1.13\pm 0.01$ (the difference mostly coming from the nodal
segment being completely occupied) which is reasonably close to
the doping.

The inset to Fig. 2d shows a comparison of both raw data and data
divided by the Fermi function at the E$_F$ crossing taken at T=36
K, well above the T$_c$.  The spectra clearly shows that the lower
band never quite reaches E$_F$ and there is in fact no FS pocket
near ($\pi/2$,$\pi/2$) point. Therefore, the `FS' segment near
($\pi/2$,$\pi/2$) in Fig. 1a is due to the finite size of the
energy integration window. We indeed find that the relative
intensity of the FS near ($\pi/2$,$\pi/2$) quickly drops in
comparison to the other as we reduce the size of the integration
window.  In SCCO the $V_{\pi\pi}$ is sufficiently large to push
the FS folded bands below E$_F$ along the zone diagonal. Recently,
there have been conflicting reports on the gap symmetry determined
by magnetic penetration depth measurements\cite{Kim,Snezhko}. If -
like other $n$-type cuprates - superconducting SCCO at this doping
level is $d$-wave\cite{Matsui1,Armitage1,Tsuei}, our result
directly shows that a `nodeless gap' can be compatible with
$d$-wave symmetry and provides a clue to the gap symmetry
controversies. To our knowledge, this is the first momentum
resolved experimental evidence for a nodeless $d$-wave gap with a
gap only near ($\pi$,0) and no low energy zone diagonal states.

\begin{figure}[t]
\centering \leavevmode \epsfxsize=8.5cm \epsfbox{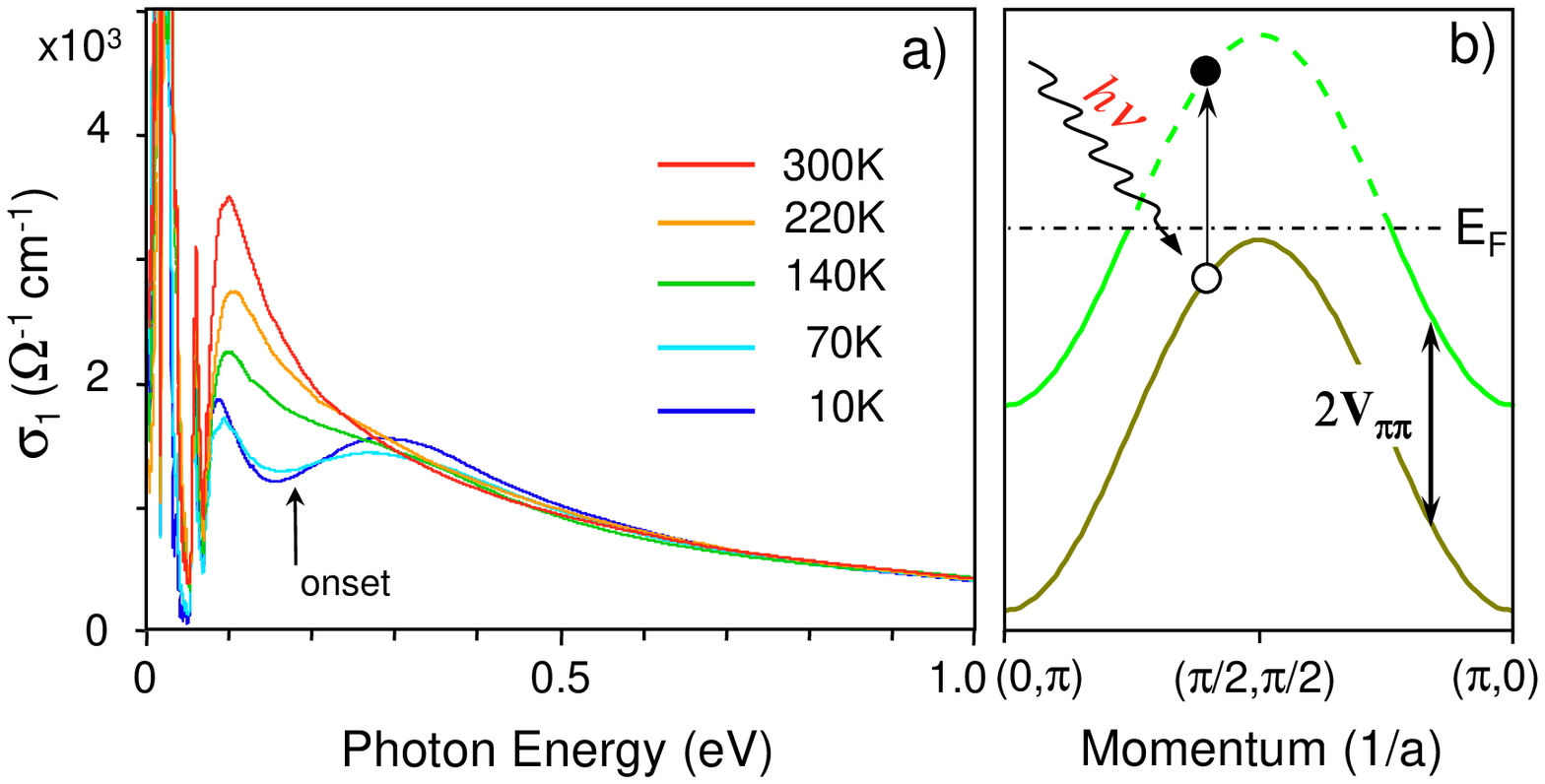}
\vspace{.0cm} \caption{Real part of the conductivity at different
temperatures obtained by reflectivity measurements. The arrow
indicates the onset of the inter-band transition. The panel on the
right is a schematic of the band structure and inter-band
transition along the AFZB.}\label{fig3}
\end{figure}

Pseudo-gap effects are also seen in optical
measurements\cite{Onose} and connection to the pseudo-gap in SDW
model has been proposed\cite{Zimmers}. However, a clear
experimental connection between the two on the same system has
previously never been obtained. As the upper band is partially
filled, inter-band transitions between the two band sheets are
possible which can be confirmed by optical measurements. Shown in
Fig. 3 is the optical conductivity at different temperatures
obtained by reflectivity measurements. In the 10K spectrum, one
can see a peak near 300 meV with an onset near 200 meV. This is
compatible with the inter-band transition inferred from the ARPES
of 2$V_{\pi\pi} = 220$ meV and allows us to interpret the two
phenomena (pseudo-gaps in ARPES and IR) in a unified picture. As
the temperature increases, the inter-band transition peak weakens,
implying weakening of $\sqrt{2}\times\sqrt{2}$ reconstruction.
Note, however, that the energy position of the peak remains more
or less the same. Similar but weaker features have been observed
in other compounds and referred to as a
pseudo-gap\cite{Zimmers,Onose}. We also note that the fact that
onset energy in $x$=0.15 NCCO is smaller than that of SCCO is
compatible with the fact that NCCO has a smaller $V_{\pi\pi}$.
Considering the fact that SCCO has lower optimal T$_c$, the
pseudo-gap appears to have an anti-correlation with
T$_c^{optimal}$.

Even though the simple SDW model explains various aspects of the
electronic structure in electron doped HTSCs, there are important
aspects with which it is not compatible. Panels (a) through (d) in
Fig. 4 show various cuts as marked in panel (f). These data are
similar to other recent ones on NCCO\cite{Matsui2}. From the fact
that the minimum splitting between the upper and lower SDW bands
appears to change with momentum, an anisotropic SDW gap was argued
for\cite{Matsui2}. However, in such a case, the two bands should
still backfold symmetrically across the AFZB unlike as observed.
Here it can been seen that while the upper band more or less ends
exactly at the AFZB, the lower one clearly overshoots the AFZB,
which is incompatible with an ordering vector ($\pi$,$\pi$).
Instead, the lower one is more reminiscent of the original
unreconstructed band. This aspect of the data has been previously
overlooked.

\begin{figure}[t]
\centering \leavevmode \epsfxsize=8.5cm \epsfbox{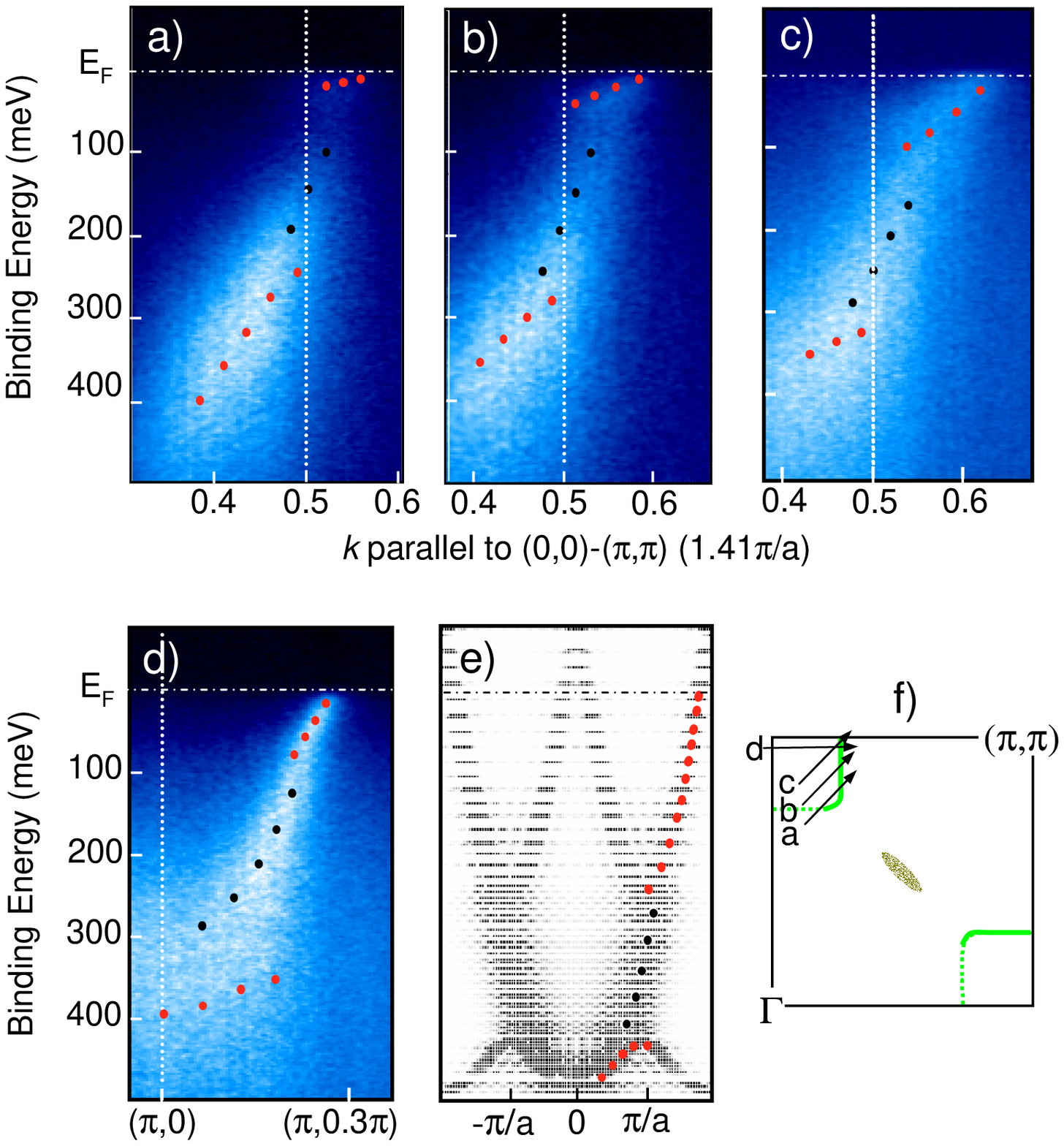}
\vspace{.0cm} \caption{ARPES data in panels (a) through (d) are
taken along the cuts marked in panel (f). (e) Eigensolutions based
on numerical model detailed in text, which mimics a short range
ordered state. Marked in the figure by black dots corresponds to
the original band and the red dots to reconstructed
one.}\label{fig4}
\end{figure}

The observed discrepancy between the data and model could be
attributed to existence of short range ordering or an
inhomogeneous magnetic state. To illustrate the effect of a short
range ordering, we consider 1D electrons that are subject to short
range periodic potentials of strength $V_{\pi\pi}$ and periodicity
$a$. Energy eigenvalue solutions are obtained for potentials with
various extents and plotted in panel (e). This simulates a
situation where electrons feel only a short range order. One notes
that it gives not only the folded bands (red dots) but also the
original band (black dots). In this interpretation, the middle
band in panels (a)-(d) represents the original band which
disperses uninterrupted through the AFZB. In fact, while recent
neutron scattering results show that a long range AF order can not
coexist with superconductivity\cite{Motoyama}, a short range AF
order exists at low temperature\cite{Yamada} in NCCO. If the short
range $\sqrt{2}\times\sqrt{2}$ ordering is indeed from AF
fluctuations, the temperature dependence of the optical spectrum
is naturally explained. At a very low temperature, the two bands
are relatively well defined and so is the inter-band transition
peak. As the temperature is raised, the AF coherence length gets
shorter and the two bands become ill defined. As a consequence,
the inter-band transition peak becomes weaker, resulting in a
filling of the `pseudo-gap'. This naturally connects the AF energy
scales with that of the pseudo-gap formation. It is important to
note in this picture that it is not the energy scale $V_{\pi\pi}$
but the ordering length that changes. Therefore, the pseudo-gap
does not close but instead fills in as the temperature increases.
Alternatively the observed ARPES spectra can be explained in terms
of an energy scale where features below about 100 meV derive from
reconstructed bands and above 100 meV from the original band.
These observations can be explained with a ($\pi$,$\pi$)
scattering which becomes ineffective above a certain energy scale
(not necessarily momentum dependent). A similar ``ineffectiveness"
may have been observed in the checker board pattern measured by
scanning tunnelling microscopy on (Na,Ca)$_2$CuO$_2$Cl$_2$ which
disappears for $\omega$ above 50 meV\cite{Hanaguri}.

Finally, another discrepancy between the data and the simple model
is in the intensity of the folded band. One can obtain the
expected relative intensities of the folded bands from the
eigensolution of the 2$\times$2 matrix equation. Even though band
folding is clearly observed as shown in Fig. 2, the intensity
appears weaker than predicted within the model. For example, the
intensity of the folded band at the $k$-point marked by a green
dot in Fig. 1a should be about 1/4 of the intensity at the cross,
but is observed to be much weaker. This observation is independent
of the photon energies or Brillouin zone, thus showing it is not
due to the matrix element effect. This aspect of the data should
be described by a more rigorous theory. Interestingly, many-body
calculations based on the Hubbard model, which show the existence
of a short range ordering, exhibit a weakened intensity of the
folded bands\cite{Tremblay}.

Authors thank T. Tohyama, A.M. Tremblay and J.H. Han for helpful
discussions and D. H. Lu for technical assistance. This work is
supported (in part) by KOSEF through CSCMR. NPA is supported the NSF
IRF Program and MaNEP. ALS and SSRL are operated by the DOE¡¯s
Office of BES.


\begin{thebibliography}{17}
\bibitem{SDW} We do not necessarily suggest spin origin. Any
$\sqrt{2}\times\sqrt{2}$ static ordering would produce similar
results.
\end{thebibliography}
\end{document}